\documentclass[aps,pre,amsmath,amssymb,longbibliography,lengthcheck,showpacs,floatfix]{revtex4-1}
\usepackage{graphicx}
\def\vec#1{\mbox{\boldmath $#1$}}
\newcommand{\vd}{\mathcal{D}}

\begin{document}

\title{Dynamical heterogeneity in a highly supercooled liquid:
Consistent calculations of correlation length, intensity, and lifetime}

\author{Hideyuki Mizuno}
 \email{h-mizuno@cheme.kyoto-u.ac.jp}
\author{Ryoichi Yamamoto}
 \email{ryoichi@cheme.kyoto-u.ac.jp}
\affiliation{Department of Chemical Engineering, Kyoto University, Kyoto 615-8510, Japan}
\affiliation{CREST, Japan Science and Technology Agency, Kawaguchi 332-0012, Japan}

\date{\today}

\begin{abstract}
We have investigated dynamical heterogeneity in a highly supercooled liquid using molecular-dynamics simulations in three dimensions.
Dynamical heterogeneity can be characterized by three quantities: correlation length $\xi_4$, intensity $\chi_4$, and lifetime $\tau_{\text{hetero}}$.
We evaluated all three quantities consistently from a single order parameter.
In a previous study (H. Mizuno and R. Yamamoto, Phys. Rev. E {\bf 82}, 030501(R) (2010)), we examined the lifetime $\tau_{\text{hetero}}(t)$ in two time intervals $t=\tau_\alpha$ and $\tau_{\text{ngp}}$, where $\tau_\alpha$ is the $\alpha$-relaxation time and $\tau_{\text{ngp}}$ is the time at which the non-Gaussian parameter of the Van Hove self-correlation function is maximized.
In the present study, in addition to the lifetime $\tau_{\text{hetero}}(t)$, we evaluated the correlation length $\xi_4(t)$ and the intensity $\chi_4(t)$ from the same order parameter used for the lifetime $\tau_{\text{hetero}}(t)$.
We found that as the temperature decreases, the lifetime $\tau_{\text{hetero}}(t)$ grows dramatically, whereas the correlation length $\xi_4(t)$ and the intensity $\chi_4(t)$ increase slowly compared to $\tau_{\text{hetero}}(t)$ or plateaus.
Furthermore, we investigated the lifetime $\tau_{\text{hetero}}(t)$ in more detail.
We examined the time-interval dependence of the lifetime $\tau_{\text{hetero}}(t)$ and found that as the time interval $t$ increases, $\tau_{\text{hetero}}(t)$ monotonically becomes longer and plateaus at the relaxation time of the two-point density correlation function.
At the large time intervals for which $\tau_{\text{hetero}}(t)$ plateaus, the heterogeneous dynamics migrate in space with a diffusion mechanism, such as the particle density.
\end{abstract}

\pacs{64.70.P-, 61.20.Lc, 61.43.Fs}

\maketitle

\section{INTRODUCTION}
As liquids are cooled toward the glass transition temperature $T_g$, a drastic slowing occurs for dynamical properties such as the structural relaxation time, the diffusion constant, and the viscosity, while only small changes are detected in static properties \cite{ediger_1996,Debenedetti_2001}.
Despite the extremely widespread use of glass in industry, the formation process and dynamic properties of this material are still poorly understood.
The goal of theoretical investigations of the glass transition is to understand the universal mechanism that gives rise to the drastic slowing of dynamical properties.
Numerous studies have attempted to explain the fundamental mechanisms of the slowing of the dynamics observed in fragile glass (i.e., the sharp increase in viscosity near the glass transition).
However, the physical mechanisms behind this slowing have not been successfully identified.

Recently, ``dynamical heterogeneities'' in glass-forming liquids have attracted much attention.
The dynamics of glass-forming liquids are not only drastically slow but also become progressively more heterogeneous upon approaching the glass transition.
Dynamical heterogeneities have been detected and visualized through simulations of soft-sphere systems \cite{muranaka_1994,hurley_1995,yamamoto1_1998,yamamoto_1998,perera_1999,cooper_2004}, hard-sphere systems \cite{doliwa_2002}, and Lennard-Jones (LJ) systems \cite{donati_1998}, and through experiments performed on colloidal dispersions using particle-tracking techniques \cite{Marcus1999,kegel_2000,weeks_2000}.
Insight into the mechanisms of dynamical heterogeneities will lead to a better understanding of the slowing of the dynamics near the glass transition.

The properties of dynamical heterogeneity can be characterized by the following three quantities: the correlation length, the intensity, and the lifetime.
In a system displaying dynamical heterogeneity, the particles can be divided into ``slow'' and ``fast'' sub-sets.
The slow and fast particles form cooperative correlated regions, and these slow and fast regions migrate in space over time.
These three quantities (the correlation length, the intensity, and the lifetime) can describe the static and dynamic properties of the slow and fast regions.
The intensity measures the average variance of the slow and fast regions, and the correlation length characterizes the spatial extent of the slow and fast regions.
The lifetime represents the time scale at which the slow and fast regions migrate in the space.

The correlation length, intensity, and lifetime can be investigated using the correlation functions of the particle dynamics.
In fact, we can evaluate the correlation length $\xi_4$ and the intensity $\chi_4$ by calculating the four-point correlation functions that correspond to the static structure factors of the particle dynamics.
Several simulations \cite{yamamoto1_1998,perera_1999,glotzer_2000,doliwa2000,lacevic_2003,berthier_2004,toninelli_2005,chandler_2006,stein_2008,karmakar_2010,karmakar2_2010,flenner_2010,flenner_2011}, experiments \cite{ediger_2000,berthier_2005,Ferrier2007,narumi_2011}, and mode-coupling theories \cite{biroli_2004,biroli_2006,szamel_2010} have estimated $\xi_4$ and $\chi_4$ using four-point correlation functions and have revealed that $\xi_4$ and $\chi_4$ increase with decreasing temperature (or an increase in the volume fraction in the case of hard-sphere systems).
Furthermore, we can quantify the lifetime $\tau_{\text{hetero}}$ using the multiple-time extensions of the four-point correlation functions (i.e., the multi-time correlation functions) that correspond to the time correlation functions of the particle dynamics.
Recent simulations have quantified $\tau_{\text{hetero}}$ using multi-time correlation functions \cite{yamamoto_1998,flenner_2004,leonard_2005,kim_2009,kim_2010}.
Various experiments, including photobleaching techniques and nuclear magnetic resonance, have also measured $\tau_{\text{hetero}}$ \cite{ediger_2000,richert_2002,wang_1999,wang_2000,rohr_1991,bohmer_1991,russell_2000}.
It was reported that $\tau_{\text{hetero}}$ increases dramatically with decreasing temperature or an increase in the volume fraction and can exceed the $\alpha$-relaxation time near the glass transition.

As mentioned above, there have been many studies on the correlation length, the intensity, and the lifetime of dynamical heterogeneity near the glass transition.
However, knowledge of and measurements relating to the lifetime are still limited.
Moreover, individual studies have been restricted to only the correlation length and the intensity or only the lifetime, and the relationship between the length and time scales of dynamical heterogeneity remains controversial despite its importance \cite{kim_2009,kim_2010}.

The aim of the present study is to examine all the three quantities (the correlation length, the intensity, and the lifetime of dynamical heterogeneity) consistently.
We performed molecular-dynamics (MD) simulations and investigated dynamical heterogeneity using the correlation functions of the particle dynamics.
In our previous study \cite{mizuno_2010}, we evaluated the lifetime $\tau_{\text{hetero}}(t)$ in two different time intervals: the $\alpha$-relaxation time $\tau_\alpha$ and the time $\tau_{\text{ngp}}$ at which the non-Gaussian parameter of the Van Hove self-correlation function is maximized.
In the present study, in addition to the lifetime $\tau_{\text{hetero}}(t)$, we quantified the correlation length $\xi_4(t)$ and the intensity $\chi_4(t)$ from the same order parameter used when calculating $\tau_{\text{hetero}}(t)$.
Furthermore, we examined the time-interval dependence of the lifetime $\tau_{\text{hetero}}(t)$ to understand the lifetime and the dynamic properties of dynamical heterogeneity in more detail.

The paper is organized as follows.
In Sec. \ref{pdynamics}, we explain the correlation functions of the particle dynamics.
We show that dynamical heterogeneity can be systematically examined using the correlation functions of the particle dynamics.
In Sec. \ref{model}, we briefly review our MD simulation and present some results from conventional density correlation functions.
In Secs. \ref{result1} and \ref{result2}, the results for dynamical heterogeneity are presented.
In Sec. \ref{result1}, we first show three quantities, the correlation length, the intensity, and the lifetime, which are consistently calculated from a single order parameter.
In Sec. \ref{result2}, we next present the time-interval dependence of the lifetime $\tau_{\text{hetero}}(t)$.
In Sec. \ref{conclusion}, we summarize our results.

\section{CORRELATION FUNCTIONS OF PARTICLE DYNAMICS} \label{pdynamics}
As we mentioned, dynamical heterogeneity can be characterized by three quantities: the correlation length, the intensity, and the lifetime.
In this section, we introduce the correlation functions of the particle dynamics and demonstrate that these three quantities can be systematically evaluated in terms of the correlation functions of the particle dynamics.

The conventional two-point correlation function $F({k},t)$ represents the correlation of local fluctuations $\delta n(\vec{k},t)$ with some order parameter, such as particle density.
The expression $\delta n(\vec{k},t)$ is the Fourier component \vec{k} of the fluctuations at time $t$, and $F({k},t) = \langle \delta n(\vec{k},t) \delta n(-\vec{k},0) \rangle$, where $k=\left| \vec{k} \right|$. When $t=0$, $S({k}) \equiv F(k,t=0)$ is the spatial correlation of $\delta n(\vec{k},0)$ (i.e., the static structure factor), and we can examine the static structure of the order parameter by the wavenumber dependence of $S({k})$. When $t>0$, $F(k,t)$ describes the particle dynamics in the time interval $[0,t]$, averaged over the initial time and space.
As the time interval $t$ increases, $F({k},t)$ decays in the stretched exponential form,
\begin{equation}
\frac{F({k},t)}{F({k},0)} \sim \exp \left( -\left( \frac{t}{\tau({k})}\right)^\beta \right),
\label{F2}
\end{equation}
where $\tau({k})$ is the relaxation time of the two-point correlation function that represents the characteristic time scale of the average particle dynamics.

\begin{figure}
\begin{center}
\includegraphics[scale=1]{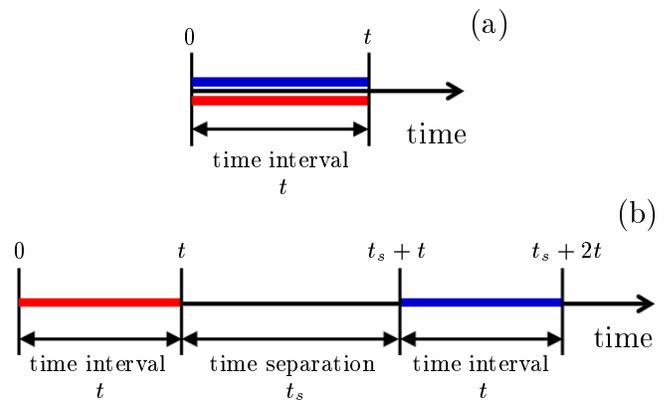}
\end{center}
\vspace*{-3mm}
\caption{(Color)
Schematic illustration of the time configuration of the correlation functions of the particle dynamics: 
(a) the spatial correlation function of the particle dynamics in the time interval $[0,t]$ and
(b) the time correlation function of the particle dynamics between the two time intervals $[0,t]$ and $[t_s+t,t_s+2t]$.
}
\label{ts}
\end{figure}

To examine the structure and motion of spatially heterogeneous dynamics, we must calculate the correlation of the local fluctuations, $\delta Q_{{k}}(\vec{q},t_0,t)$, in the particle dynamics.
The expression $\delta Q_{{k}}(\vec{q},t_0,t)$ is the Fourier component $\vec{q}$ of the fluctuations in the particle dynamics that are associated with a microscopic wavenumber ${k}$ in the time interval $[t_0,t_0+t]$.
$F({k},t)$ is equal to $Q_{{k}}(\vec{q},t_0,t)$ averaged over the initial time $t_0$ and space, i.e., $F({k},t) \sim \langle Q_{{k}}(\vec{q},t_0,t) \rangle$. The correlation function defined by
\begin{equation}
\begin{aligned}
S_{4,{k}}({q},t)=
\langle \delta Q_{{k}}(\vec{q},0,t) \delta Q_{{k}}(-\vec{q},0,t) \rangle,
\end{aligned} \label{K4}
\end{equation}
represents the spatial correlation of the particle dynamics in the time interval $[0,t]$.
The time configuration of $S_{4,{k}}({q},t)$ is schematically illustrated in Fig. \ref{ts}(a). $S_{4,{k}}({q},t)$ is the four-point correlation function.
We can examine the structure of the particle dynamics by the wavenumber dependence of $S_{4,{k}}({q},t)$.
At small wavenumbers of $q$, $S_{4,k}(q,t)$ can be approximated by the simple Ornstein-Zernike (OZ) form \cite{yamamoto1_1998,lacevic_2003},
\begin{equation}
S_{4,k}(q,t) = \frac{\chi_{4,k}(t)}{1 + q^2 \xi_{4,k}(t)^2}, \label{OZ}
\end{equation}
where $\xi_{4,k}(t)$ is the correlation length of the heterogeneous dynamics and $\chi_{4,k}(t)$ is the intensity of the heterogeneous dynamics, which is the long-wavelength limit of $S_{4,k}(q,t)$, i.e., $\chi_{4,k}(t) = \lim_{q \rightarrow 0} S_{4,k}(q,t)$.
Note that $\chi_{4,k}(t)$ is termed the four-point dynamical susceptibility.
According to previous studies \cite{karmakar_2010,karmakar2_2010,flenner_2010,flenner_2011}, to quantify $\xi_{4,k}(t)$ and $\chi_{4,k}(t)$ accurately, we need to use a large system or determine $\chi_{4,k}(t)$ through other means (not OZ fitting).
In their work \cite{flenner_2010,flenner_2011}, E. Flenner et al. claimed that accurate determinations of $\xi_{4,k}(t)$ and $\chi_{4,k}(t)$ can be made by fitting $S_{4,k}(q,t)$ to the OZ form in the range of $q \xi_{4,k}(t) < 1.5$.
They used a large system composed of $8 \times 10^4$ particles to obtain accurate fits for the OZ form.
In the present study, we used a large system with $10^5$ particles to fit $S_{4,k}(q,t)$ to the OZ form accurately and obtain the values of $\xi_{4,k}(t)$ and $\chi_{4,k}(t)$.

Furthermore, the time correlation function defined by
\begin{equation}
\begin{aligned}
F_{4,{k}}({q},t_s,t)=
\langle \delta Q_{{k}}(\vec{q},t_s+t,t) \delta Q_{{k}}(-\vec{q},0,t) \rangle,
\end{aligned} \label{F4}
\end{equation}
represents the correlation of the particle dynamics between the two time intervals $[0,t]$ and $[t_s+t,t_s+2t]$.
The value $t_s$ is the time separation between the two time intervals $[0,t]$ and $[t_s+t,t_s+2t]$.
The time configuration of $F_{4,{k}}({q},t_s,t)$ is schematically illustrated in Fig. \ref{ts}(b). $F_{4,{k}}({q},t_s,t)$ is the multiple-time extension of the four-point correlation function \cite{kim_2009,kim_2010}.
As the time separation $t_s$ increases, $F_{4,{k}}({q},t_s,t)$ with fixed $t$ decays in the stretched exponential form,
\begin{equation}
\begin{aligned}
\frac{F_{4,{k}}(q,t_s,t)}{F_{4,{k}}({q},0,t)}
\sim \exp\left( - \left( \frac{t_s}{\tau_{4,{k}}({q},t)} \right)^c \right),
\end{aligned} \label{F42}
\end{equation}
where $\tau_{4,{k}}({q},t)$ is the relaxation time of the correlation of the particle dynamics.
We determined the lifetime $\tau_{\text{hetero}}(t)$ of the heterogeneous dynamics as $\tau_{4,{k}}({q},t)$ at $q=0.38$.
We need longer trajectories of the simulations to quantify $\tau_{\text{hetero}}(t)$ than to quantify $\xi_{4,k}(t)$ and $\chi_{4,k}(t)$.
In the present study, we used a smaller system with $10^4$ particles to calculate $\tau_{\text{hetero}}(t)$.
As explained above, we can systematically evaluate the correlation length, the intensity, and the lifetime of dynamical heterogeneity by calculating the correlation functions of the particle dynamics.

\section{SIMULATION MODEL AND RESULTS FROM THE DENSITY CORRELATION FUNCTIONS} \label{model}
\subsection{Simulation model}
We performed MD simulations in three dimensions on binary mixtures of two different atomic species, 1 and 2, with a cube of constant volume $V$ as the basic cell, surrounded by periodic boundary image cells.
The particles interacted via their soft-sphere potentials, $v_{a b}(r)= \epsilon (\sigma_{a b}/r)^{12}$; where r is the distance between two particles, $\sigma_{a b} = (\sigma_a + \sigma_b)/2$, and $a,b \in 1,2$. The interaction was truncated at $r=3 \sigma_{a b}$. The mass ratio was $m_2/m_1=2$, and the diameter ratio was $\sigma_2/\sigma_1=1.2$.
This diameter ratio avoided system crystallization and ensured that an amorphous supercooled state formed at low temperatures \cite{miyagawa_1991}.
As mentioned in Sec. \ref{pdynamics}, we used two systems: a small system with $N_1 = N_2 =5 \times 10^3$ ($N=N_1+N_2=10^4$) particles and a large system with $N_1 = N_2 =5 \times 10^4$ ($N=N_1+N_2=10^5$) particles.
The large system was used to quantify the correlation length $\xi_{4,k}(t)$ and the intensity $\chi_{4,k}(t)$, and the small system was used to quantify the lifetime $\tau_{\text{hetero}}(t)$.
In the present paper, the following dimensionless units are used: length, $\sigma_1$; temperature, $\epsilon/k_B$; and time, $\tau_0 = (m_1 \sigma_1^2/\epsilon)^{1/2}$. The particle density was fixed at the high value of $\rho = N/V = 0.8$.
The system lengths were $L=V^{1/3} = 23.2$ and $50.0$ for the small and large systems, respectively.
Simulations were performed at $T=0.772,\ 0.473,\ 0.352,\ 0.306,\ 0.289,\ 0.267$, and $0.253$.
Note that the freezing point of the corresponding one-component model is approximately $T=0.772\ (\Gamma_{\text{eff}}=1.15)$ \cite{miyagawa_1991}.
Here, $\Gamma_{\text{eff}}$ is the effective density, which is a single parameter characterizing this model.
At $T=0.253\ (\Gamma_{\text{eff}}=1.52)$, the system is in a highly supercooled state.
We used the leapfrog algorithm with time steps of $0.005$ when integrating the Newtonian equation of motion.
At each temperature, the system was carefully equilibrated under the canonical condition so that no appreciable aging effect was detected for various quantities, including the pressure and the density correlation function.
Once equilibrium was established, data were taken under the microcanonical condition.
The length of the data collection runs was at least 100 times the $\alpha$-relaxation time, $\tau_\alpha$, for the small system and 10 times $\tau_\alpha$ for the large system.
Information regarding this model, such as the static structure factor, the intermediate scattering function, and the mean square displacement, can be found in previous works \cite{yamamoto1_1998,kim_2000}.

\subsection{Single-particle and collective-particle diffusive motion} \label{sincol}
Before showing the results for dynamical heterogeneity in supercooled liquids, we present some results from an investigation of the average particle dynamics using conventional density correlation functions.
Let us consider the density correlation functions $F_{sa}(k,t)$ and $F_a(k,t)\ (a \in 1,2)$, defined by
\begin{equation}
\begin{aligned}
& F_{sa}(k,t) = \left< \frac{1}{N_a} \sum_{j=1}^{N_a} \delta n_{aj}(\vec{k},t) \delta n_{aj}(-\vec{k},0) \right>, \\
& F_a(k,t) = \langle \delta n_a(\vec{k},t) \delta n_a(-\vec{k},0) \rangle, \label {dc}
\end{aligned}
\end{equation}
where $\delta n_{aj}(\vec{k},t) = \exp[ -i \vec{k} \cdot \vec{r}_{aj}(t)]$ is the Fourier component $\vec{k}$ of the tagged particle density fluctuations of particle species $a$, and $\delta n_a(\vec{k}, t) = \sum^{N_a}_{j=1} \exp[ -i \vec{k} \cdot \vec{r}_{aj}(t)]$ is the Fourier component $\vec{k}$ of the density fluctuations of particle species $a$. The terms $F_{sa}(k,t)$ and $F_a(k,t)$ describe the single-particle and collective-particle motion, respectively \cite{simpleliquid}.
We calculated $F_{sa}(k,t)$ and $F_a(k,t)$ for a wide range of wavenumbers, $k=0.35 - 40$.
As seen in Eq. (\ref{F2}), $F_{sa}(k,t)$ and $F_a(k,t)$ decay in the stretched exponential form,
\begin{equation}
\begin{aligned}
& \frac{F_{sa}(k,t)}{F_{sa}(k,0)} = \exp \left( -\left( \frac{t}{\tau_{sa}(k)} \right)^{\beta_s} \right),\\
& \frac{F_a(k,t)}{F_a(k,0)} = \exp \left( -\left( \frac{t}{\tau_{ca}(k)} \right)^{\beta_c} \right),
\end{aligned}
\end{equation}
where $\tau_{sa}(k)$ and $\tau_{ca}(k)$ are the wavenumber-dependent relaxation times of $F_{sa}(k,t)$ and $F_a(k,t)$, respectively.

\begin{figure}
\begin{center}
\includegraphics[scale=1]{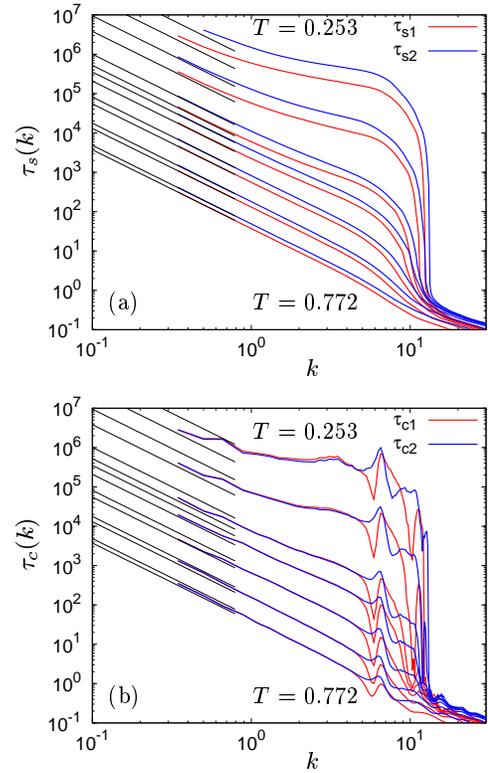}
\end{center}
\vspace*{-3mm}
\caption{(Color) 
The wavenumber dependence of (a) $\tau_{sa}(k)$ and (b) $\tau_{ca}(k)$ for particle species $a=1$ and $2$.
Temperatures are $0.772 - 0.253$ from the lowest curve to the highest.
The black line is $\tau_{sa}(k) = D_{sa}^{-1} k^{-2}$, where $D_{sa}$ is the diffusion constant of the single-particle motion calculated from the mean square displacement.
}
\label{tauk}
\end{figure}

In Fig. \ref{tauk}, $\tau_{sa}(k)$ and $\tau_{ca}(k)$ are plotted for particle species 1 and 2 as functions of the wavenumber $k$.
In Fig. \ref{tauk}(a), $\tau_{sa}(k)$ approaches $\tau_{sa}(k) = D_{sa}^{-1} k^{-2}$ at small wavenumbers $k$, where $D_{sa}$ is the diffusion constant of the single-particle motion of particle species $a$. The term $D_{sa}$ is calculated by $D_{sa} = \lim_{t \rightarrow \infty} \langle [\Delta\vec{r}_a(t)]^2 \rangle/6t$; where $\langle [\Delta\vec{r}_a(t)]^2 \rangle$ is the mean square displacement of particle species $a$; $\langle [\Delta\vec{r}_a(t)]^2 \rangle = \langle \sum_{j=1}^{N_a} [\Delta\vec{r}_{aj}(t)]^2/N_a \rangle$; $\Delta\vec{r}_{aj}(t) = \vec{r}_{aj}(t)-\vec{r}_{aj}(0)$.
The diffusion constant of particle species 1 is larger than that of particle species 2 at every temperature.
However, $\tau_{ca}(k)$ also approaches $\tau_{ca}(k) \sim k^{-2}$ at small $k$ in Fig. \ref{tauk}(b).
This behavior indicates that the collective-particle motion is also diffusive at large length scales, and we can define the diffusion constant of the collective-particle motion $D_a$ as
\begin{equation}
D_a=\lim_{k \to 0} \tau_{ca}^{-1}(k) k^{-2}.
\end{equation}
We can see that $D_1$ and $D_2$ are almost identical between $D_{s1}$ and $D_{s2}$ in Fig. \ref{tauk}(b).

\begin{figure}
\begin{center}
\includegraphics[scale=1]{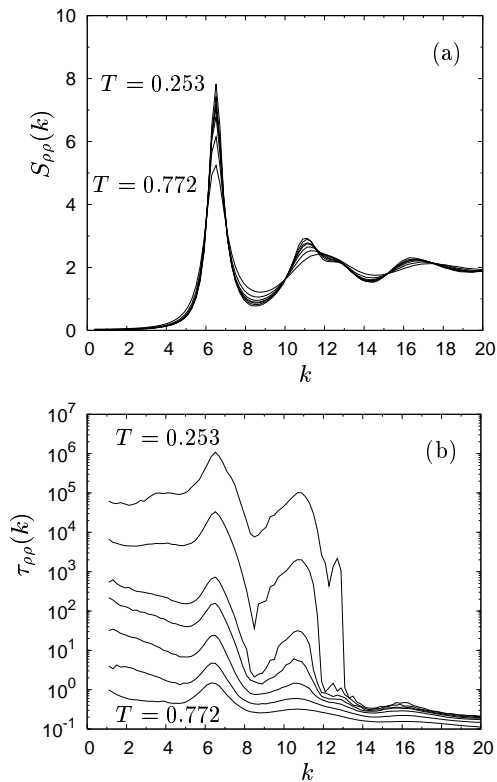}
\end{center}
\vspace*{-3mm}
\caption{
The wavenumber dependence of (a) $S_{\rho\rho}(k)$ and (b) $\tau_{\rho\rho}(k)$.
The temperatures are $0.772 - 0.253$ from the lowest curve to the highest.
}
\label{rhorho}
\end{figure}

\subsection{Correlation between the static structure factor and the relaxation time}
Next, we consider a density variable defined by
\begin{equation}
\rho_{\text{eff}}(\vec{k},t) = \sigma_1^3 n_1(\vec{k},t) + \sigma_2^3 n_2(\vec{k},t),
\end{equation}
that represents the degree of particle packing and is the effective one-component density of our binary mixture \cite{yamamoto1_1998}.
We calculated the static structure factor $S_{\rho\rho}(k)$ and the time correlation function $F_{\rho\rho}(k,t)$, defined by
\begin{equation}
\begin{aligned}
& S_{\rho\rho}(k) = \frac{1}{N} \langle \delta \rho_{\text{eff}}(\vec{k},0) \delta \rho_{\text{eff}}(-\vec{k},0) \rangle ,\\
& F_{\rho\rho}(k,t) = \langle \delta \rho_{\text{eff}}(\vec{k},t) \delta \rho_{\text{eff}}(-\vec{k},0) \rangle.
\end{aligned}
\end{equation}
The terms $S_{\rho\rho}(k)$ and $F_{\rho\rho}(k,t)$ represent the static structure and the average particle dynamics of the effective one-component fluids, respectively.
Note that $F_{\rho\rho}(k,t)$ decays in the stretched exponential form with the relaxation time $\tau_{\rho\rho}(k)$, as in Eq. (\ref{F2}).

Figure \ref{rhorho} shows the wavenumber dependence of $S_{\rho\rho}(k)$ in (a) and $\tau_{\rho\rho}(k)$ in (b).
We can see that $S_{\rho\rho}(k)$ and $\tau_{\rho\rho}(k)$ are maximized and minimized at almost the same wavenumbers $k$. For example, both $S_{\rho\rho}(k)$ and $\tau_{\rho\rho}(k)$ have first-peak values around $k=2\pi$.
Thus, there is a correlation between $S_{\rho\rho}(k)$ and $\tau_{\rho\rho}(k)$, i.e., between the static structure and the particle dynamics.
A similar correlation was found in the Lennard-Jones (LJ) model \cite{degennes1}, water \cite{degennes2}, and the polymer model \cite{degennes3}.
According to Ref. \cite{degennes3}, $\tau_{\rho\rho}(k)$ is modulated by $S_{\rho\rho}(k)$, and this modulation can be understood as a consequence of ``de Gennes narrowing'' \cite{degennes0}.

\begin{figure}
\begin{center}
\includegraphics[scale=1]{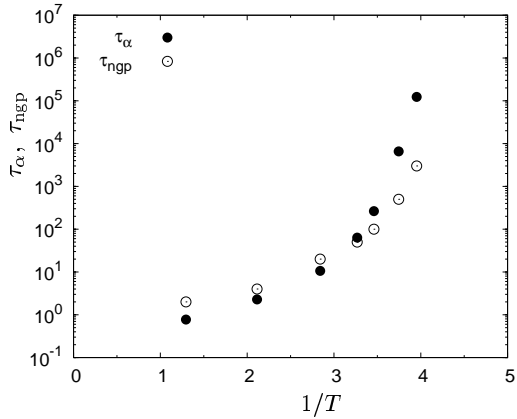}
\end{center}
\vspace*{-3mm}
\caption{
Time intervals $\tau_\alpha$ and \ $\tau_{\text{ngp}}$ versus the inverse temperature $1/T$.
We use these two time intervals to define the local dynamics.
}
\label{taua}
\end{figure}

\section{RESULTS I: CORRELATION LENGTH, INTENSITY, AND LIFETIME} \label{result1}
In this section, we present the results of three quantities: the correlation length, the intensity, and the lifetime of dynamical heterogeneity.
We consistently evaluated these three quantities using a single order parameter representing the particle dynamics and its correlation functions.
In our previous study \cite{mizuno_2010}, we evaluated the lifetime $\tau_{\text{hetero}}(t)$.
For this study, in addition to the lifetime $\tau_{\text{hetero}}(t)$, we quantified the correlation length $\xi_4(t)$ and the intensity $\chi_4(t)$ from the same order parameter used when calculating $\tau_{\text{hetero}}(t)$.
As in our previous study \cite{mizuno_2010}, we used two time intervals, $\tau_\alpha$ and $\tau_{\text{ngp}}$, to define the local dynamics.
The value $\tau_\alpha$ is the $\alpha$-relaxation time defined by $F_{s1}(k_m,\tau_{\alpha}) = e^{-1}$, where $F_{s1}(k,t)$ is the self-part of the density time correlation function for particle species 1 as defined in Eq. (\ref{dc}) and $k_m = 2\pi$ is the first-peak wavenumber of the static structure factor.
The value $\tau_{\text{ngp}}$ is the time at which the non-Gaussian parameter $\alpha_2(t)$ \cite{rahman_1964} of the Van Hove self-correlation function defined as $\alpha_2(t) = 3\langle [\Delta\vec{r}_1(t)]^4 \rangle/5\langle [\Delta\vec{r}_1(t)]^2 \rangle^2-1$ is maximized.
In Fig. \ref{taua}, we show $\tau_\alpha$ and $\tau_{\text{ngp}}$ as functions of the inverse temperature $1/T$.
At $T=0.306$, $\tau_\alpha \simeq \tau_{\text{ngp}}$, and $\tau_\alpha$ grows exponentially larger than $\tau_{\text{ngp}}$ with decreasing temperature at $T<0.306$.
This trend agrees with other simulation results for LJ systems \cite{kob_1995_1,kob_1995_2}.

\subsection{The structure of the heterogeneous dynamics}
\begin{figure}
\begin{center}
\includegraphics[scale=1]{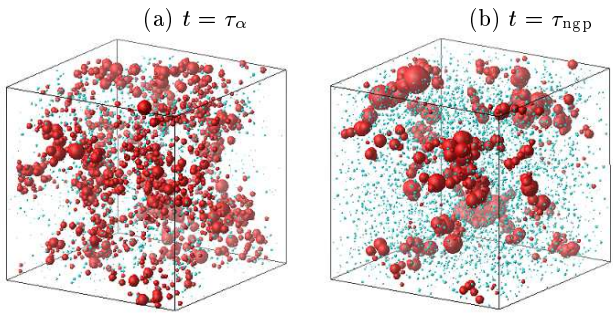}
\end{center}
\vspace*{-3mm}
\caption{(Color) 
The distribution of the particle mobility $a_{1j}^2(t_0,t)$ for particle species 1.
The time intervals are (a) $[t_0,t_0+\tau_\alpha] (t=\tau_\alpha)$ and (b) $[t_0,t_0+\tau_\text{ngp}] (t=\tau_\text{ngp})$.
The temperature is $0.253$.
The radii of the spheres are $a_{1j}^2(t_0,t)$, and the centers are at $\vec{R}_{1j}(t_0,t)$.
The red and blue spheres represent $a_{1j}^2(t_0,t) \ge 1$ (mobile particles) and $a_{1j}^2(t_0,t)<1$ (immobile particles), respectively.
}
\label{hetero1}
\end{figure}

We first examined the structure of the heterogeneous dynamics.
We calculated the displacement of each particle of species 1 in the time interval $[t_0, t_0 + t]$; $\Delta \vec{r}_{1j}(t_0,t) =\vec{r}_{1j}(t_0+t)-\vec{r}_{1j}(t_0) \ (j=1,2,...,N_1)$, and the particle mobility $a_{1j}^2(t_0,t)$ of each particle was defined as
\begin{equation}
a_{1j}^2(t_0,t) = \frac {[\Delta \vec{r}_{1j}(t_0,t)]^2}{\langle [\Delta \vec{r}_{1j}(t_0,t)]^2 \rangle}.
\end{equation}
In Fig. \ref{hetero1}, we show the spatial distribution of the particle mobility at $T=0.253$, which is the lowest temperature in our simulations.
In the figure, the particles are drawn as spheres with radii $a_{1j}^2(t_0,t)$ located at
\begin{equation}
\vec{R}_{1j}(t_0,t) = \frac{1}{2}[\vec{r}_{1j}(t_0) + \vec{r}_{1j}(t_0+t)].
\end{equation}
Notice that $a_{1j}^2(t_0,t)\ge 1$ ($a_{1j}^2(t_0,t)<1$) means that the particle $j$ moves more (less) than the mean value of the single-particle displacement, i.e., particle $j$ is mobile (immobile).
In Fig. \ref{hetero1}, the red (blue) spheres represent $a_j^2(t_0,t)\ge 1$ ($a_j^2(t_0,t)<1$).
In the figure, the two heterogeneity structures in $\tau_\alpha$ and $\tau_{\text{ngp}}$ are both significant but differ considerably.
There are many more red spheres in \ref{hetero1}(a) than in \ref{hetero1}(b), but there are much larger red spheres in \ref{hetero1}(b) than in \ref{hetero1}(a).
This pattern means that many mobile particles contribute to the heterogeneity in the time interval $\tau_\alpha$, but in the time interval $\tau_\text{ngp}$, relatively few particles are mobile and contribute to the heterogeneity.

\begin{figure}
\begin{center}
\includegraphics[scale=1]{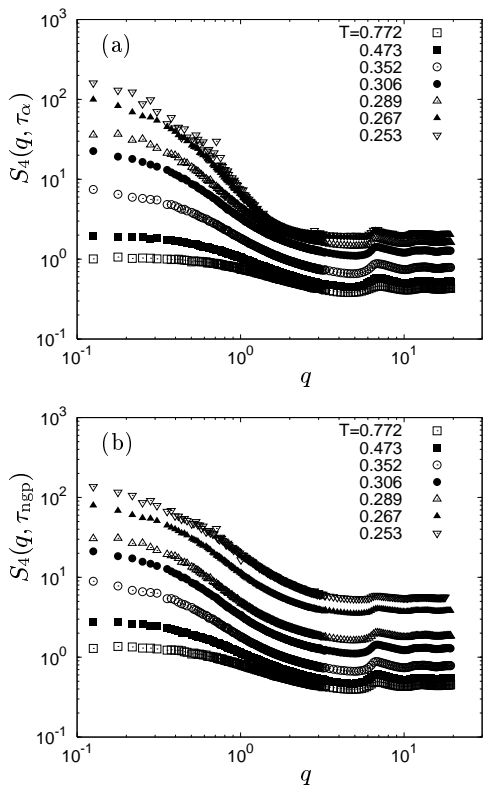}
\end{center}
\vspace*{-3mm}
\caption{
The wavenumber dependence of $S_{4}(q,t)$ for particle species 1 at $T=0.772-0.253$.
The time interval $t$ is $\tau_\alpha$ in (a) and $\tau_{\text{ngp}}$ in (b).
Note that $S_{4}(q,t)$ was calculated with a larger system, $N = 10^5$.
}
\label{sdq}
\end{figure}

We also calculated the spatial correlation function of the particle dynamics expressed in Eq. (\ref{K4}).
We considered the local fluctuations in the particle mobility of particle species 1 defined by
\begin{equation}
\delta \hat{\vd}_1(\vec{r},t_0,t) = \sum_{j=1}^{N_1} \delta a_{1j}^2(t_0,t) \delta( \vec{r} - \vec{R}_{1j}(t_0,t) ),
\end{equation}
or the Fourier component $\vec{q}$ of $\delta \hat{\vd}_1(\vec{r},t_0,t)$,
\begin{equation}
\delta \vd_1(\vec{q},t_0,t) = \sum_{j=1}^{N_1} \delta a_{1j}^2(t_0,t) \exp[-i\vec{q} \cdot \vec{R}_{1j}(t_0,t) ], \label{dop}
\end{equation}
where $\delta a_{1j}^2(t_0,t) = a_{1j}^2(t_0,t) - 1$ is the fluctuation of the particle mobility of the particle $j$.
The order parameter $\delta \vd_1(\vec{q},t_0,t)$ represents the local fluctuations in the particle dynamics in the time interval $[t_0,t_0+t]$.
Note that the same order parameter as that of $\delta \vd_1(\vec{q},t_0,t)$ is used to evaluate the lifetime in our previous study \cite{mizuno_2010}.
We used $\delta \vd_1(\vec{q},t_0,t)$ as $\delta Q_{{k}}(\vec{q},t_0,t)$ in Eq. (\ref{K4}), and the correlation function defined by
\begin{equation}
S_{4}(q,t) = \frac{1}{N_1} \langle \delta \vd_1(\vec{q},0,t) \delta \vd_1(-\vec{q},0,t) \rangle, \label{sd4eq}
\end{equation}
corresponds to $S_{4,{k}}({q},t)$. The term $S_{4}(q,t)$ represents the spatial correlation of the particle dynamics in the time interval $[0,t]$.
The time configuration of $S_{4}(q,t)$ is schematically illustrated in Fig. \ref{ts}(a).
We were able to examine the structure of the heterogeneous dynamics using the wavenumber dependence of $S_{4}(q,t)$.

Figure \ref{sdq} shows the wavenumber dependence of $S_{4}(q,t)$ of particle species 1 for $t=\tau_\alpha$ and $\tau_{\text{ngp}}$.
We calculated $S_{4}(q,t)$ using a larger system with $N = 10^5$ to quantify the correlation length and the intensity accurately.
At small wavenumbers of $q$ (long-distance scales), the correlations in $\tau_\alpha$ and $\tau_{\text{ngp}}$ both become large with decreasing temperature in a similar manner.
However, at large wavenumbers of $q$ (short-distance scales), $S_{4}(q,\tau_{\text{ngp}})$ grows larger than $S_{4}(q,{\tau_\alpha})$ at low temperatures, which reflects that more highly mobile particles exist in the time interval $\tau_\text{ngp}$, as can be seen in the visualization shown in Fig. \ref{hetero1}.

\begin{figure}
\begin{center}
\includegraphics[scale=1]{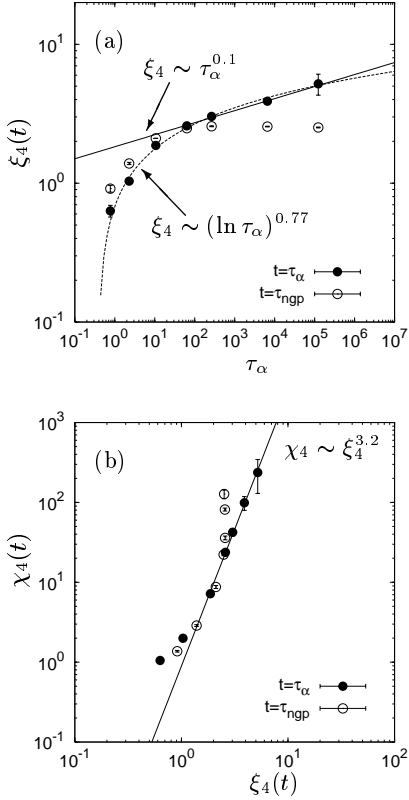}
\end{center}
\vspace*{-3mm}
\caption{
(a) The correlation length $\xi_4(t)$ versus $\tau_\alpha$.
(b) The intensity $\chi_{4}(t)$ versus the correlation length $\xi_4(t)$.
The time intervals are $t=\tau_\alpha$ and $\tau_{\text{ngp}}$.
The straight lines are power law fits: $\xi_4(\tau_\alpha) \sim \tau_\alpha^{0.1 \pm 0.01}$ and $\chi_4(\tau_\alpha) \sim \xi_4(\tau_\alpha)^{3.2 \pm 0.1}$.
The dashed curve is a fit to $\xi_4(\tau_\alpha) \sim (\ln {\tau_\alpha})^{0.77 \pm 0.12}$.
}
\label{xi}
\end{figure}

\begin{figure}
\begin{center}
\includegraphics[scale=1]{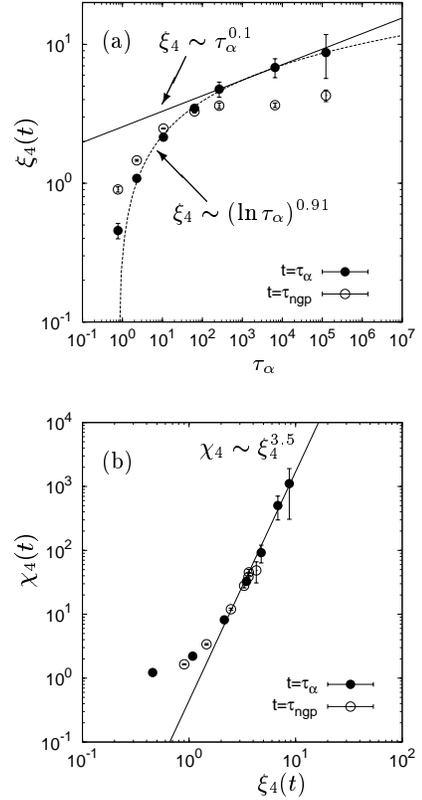}
\end{center}
\vspace*{-3mm}
\caption{
The correlation length $\xi_4(t)$ and the intensity $\chi_{4}(t)$ calculated by using $\delta \hat{\vd}_{\log 1}(\vec{r},t_0,t)$, i.e., the distribution of the logarithm of the square particle displacements.
(a) $\xi_4(t)$ versus $\tau_\alpha$.
(b) $\chi_{4}(t)$ versus $\xi_4(t)$.
The time intervals are $t=\tau_\alpha$ and $\tau_{\text{ngp}}$.
The straight lines are power law fits: $\xi_4(\tau_\alpha) \sim \tau_\alpha^{0.1 \pm 0.01}$ and $\chi_4(\tau_\alpha) \sim \xi_4(\tau_\alpha)^{3.5 \pm 0.2}$.
The dashed curve is a fit to $\xi_4(\tau_\alpha) \sim (\ln {\tau_\alpha})^{0.91 \pm 0.12}$.
}
\label{xilog}
\end{figure}

\subsection{The correlation length, the intensity, and the lifetime of the heterogeneous dynamics}
We next quantified the correlation length and the intensity of the heterogeneous dynamics in $\tau_\alpha$ and $\tau_{\text{ngp}}$.
As mentioned for Eq. (\ref{OZ}), at small wavenumbers of $q$, $S_{4}(q,t)$ can be approximated by the simple OZ form,
\begin{equation}
S_{4}(q,t) = \frac{\chi_{4}(t)}{1 + q^2 \xi_{4}(t)^2}, \label{OZd}
\end{equation}
where $\xi_{4}(t)$ is the correlation length and $\chi_{4}(t)$ is the intensity.
The values $\xi_{4}(t)$ and $\chi_{4}(t)$ correspond to $\xi_{4,k}(t)$ and $\chi_{4,k}(t)$ in Eq. (\ref{OZ}), respectively.
To obtain accurate values of $\xi_{4}(t)$ and $\chi_{4}(t)$, $S_{4}(q,t)$ was carefully fitted to the OZ form in the range of $q \xi_{4}(t) < 1.5$ \cite{flenner_2010,flenner_2011}.

Figure \ref{xi} shows $\xi_{4}(t)$ versus $\tau_\alpha$ in \ref{xi}(a) and $\chi_{4}(t)$ versus $\xi_{4}(t)$ in \ref{xi}(b) for $t=\tau_\alpha$ and $\tau_{\text{ngp}}$.
For the time interval $t=\tau_\alpha$, we examined the scaling relationships between $\tau_\alpha$, $\xi_{4}(\tau_\alpha)$, and $\chi_{4}(\tau_\alpha)$.
As in Fig. \ref{xi}(a), we obtained a power law fit $\xi_{4}(\tau_\alpha) \sim \tau_\alpha^{1/z}$ with $1/z=0.1$.
The scaling exponent $0.1$ is very small.
We also found that a fit to $\xi_4(\tau_\alpha) \sim (\ln {\tau_\alpha})^{1/\theta}$ with $1/\theta = 0.77$ provides a better description of the data over a larger range of $\tau_\alpha$, so there is a slower-than-power law increase of $\xi_4(\tau_\alpha)$ with $\tau_\alpha$.
Therefore, $\xi_{4}(\tau_\alpha)$ increases much more slowly compared to $\tau_\alpha$ as the temperature decreases.
This result is consistent with the most recent results \cite{flenner_2010,flenner_2011}.
Furthermore, we obtained the scaling relationship $\chi_{4}(\tau_\alpha) \sim \xi_4(\tau_\alpha)^{2-\eta}$ with $2-\eta=3.2$ in Fig. \ref{xi}(b).
Our scaling exponent $3.2$ agrees well with Ref. \cite{flenner_2010}.

On the other hand, when the time interval $t$ is $\tau_{\text{ngp}}$, $\xi_4(\tau_{\text{ngp}})$ reaches a value near $\tau_\alpha \simeq 250$ and then plateaus as $\tau_\alpha$ increases in Fig. \ref{xi}(a), while $\chi_4(\tau_{\text{ngp}})$ monotonically gets large in Fig. \ref{xi}(b).
Here, we remark that even in the range where $\xi_4(\tau_{\text{ngp}})$ plateaus, $\chi_4(\tau_{\text{ngp}})$ continues to increase with $\tau_\alpha$.
This behavior of $\xi_4(\tau_{\text{ngp}})$ and $\chi_4(\tau_{\text{ngp}})$ may be due to the choice of the order parameter $\delta \hat{\vd}_{1}(\vec{r},t_0,t)$.
$\delta \hat{\vd}_{1}(\vec{r},t_0,t)$ represents the distribution of the square particle displacements, so the mobile particles with large displacements contribute to the structure factor $S_{4}(q,t)$ much more than the immobile particles.
At $t=\tau_{\text{ngp}}$ and at low temperatures, our simulation data show that only a small number of particles tend to have very large displacements compared to the other mostly immobile particles, which can be also seen in Fig. \ref{hetero1}.
In this situation, the intensity $\chi_4(\tau_{\text{ngp}})$ can increase from the contribution of particles with very large displacements, whereas the correlation length $\xi_4(\tau_{\text{ngp}})$ cannot increase because the number of mobile particles involved in the correlated regions decreases.

So, we used an another order parameter $\delta \hat{\vd}_{\log 1}(\vec{r},t_0,t)$ defined as
\begin{equation}
\delta \hat{\vd}_{\log 1}(\vec{r},t_0,t) = \sum_{j=1}^{N_1} \ln a_{1j}^2(t_0,t) \delta( \vec{r} - \vec{R}_{1j}(t_0,t) ).
\end{equation}
The order parameter $\delta \hat{\vd}_{\log 1}(\vec{r},t_0,t)$ represents the distribution of the logarithm of the square particle displacements.
We quantified $\xi_4(t)$ and $\chi_4(t)$ in terms of the structure factor of $\delta \hat{\vd}_{\log 1}(\vec{r},t_0,t)$.
Notice that in this case, the immobile particles with small displacements contribute to the structure factor to the same extent as the mobile particles because of the logarithm operation.
Figure \ref{xilog} shows $\xi_4(t)$ and $\chi_4(t)$ calculated by $\delta \hat{\vd}_{\log 1}(\vec{r},t_0,t)$.
At the time interval $t=\tau_\alpha$, the scaling relationships between $\tau_\alpha$, $\xi_4(\tau_\alpha)$ and $\chi_4(\tau_\alpha)$ are almost same as those examined by $\delta \hat{\vd}_{1}(\vec{r},t_0,t)$.
On the other hand, at the time interval $t=\tau_{\text{ngp}}$, both $\xi_4(\tau_{\text{ngp}})$ and $\chi_4(\tau_{\text{ngp}})$ reach values near $\tau_\alpha \simeq 250$ and then plateau as $\tau_\alpha$ increases, i.e., both $\xi_4(\tau_{\text{ngp}})$ and $\chi_4(\tau_{\text{ngp}})$ show the plateau of the heterogeneity.
This result indicates that the behavior that $\chi_4(\tau_{\text{ngp}})$ continues to increase in Fig. \ref{xi}(a) is due to the choice of the order parameter $\delta \hat{\vd}_{1}(\vec{r},t_0,t)$.
For the confirmation purpose, we also calculated the lifetime $\tau_{\text{hetero}}(t)$ by using $\delta \hat{\vd}_{\log 1}(\vec{r},t_0,t)$ and checked that the lifetimes of both $\delta \hat{\vd}_{1}(\vec{r},t_0,t)$ and $\delta \hat{\vd}_{\log 1}(\vec{r},t_0,t)$ have almost same values and behave in the same manner with decreasing temperature.

In our previous study, we already determined the scaling relationships between the lifetime $\tau_{\text{hetero}}(t)$ and $\tau_\alpha$, i.e., $\tau_{\text{hetero}}(\tau_\alpha) \sim \tau_\alpha^{1.08}$ and $\tau_{\text{hetero}}(\tau_{\text{ngp}}) \sim \tau_\alpha^{0.91}$.
At the time interval $t=\tau_\alpha$, we can determine the scaling relationship between the correlation length and the lifetime:
\begin{equation}
\tau_{\text{hetero}}(\tau_\alpha) \sim \xi_4(\tau_\alpha)^{10.8},
\end{equation}
or
\begin{equation}
\tau_{\text{hetero}}(\tau_\alpha) \sim \exp(k \xi_4(\tau_\alpha)^{1.3}).
\end{equation}
The scaling exponent $10.8$ of the power law scaling is very large, and there is an exponential growth of $\tau_\text{hetero}(\tau_\alpha)$ with $\xi_4(\tau_\alpha)$.
Furthermore, at the time interval $t=\tau_\text{ngp}$, although the correlation length $\xi_4(\tau_{\text{ngp}})$ plateaus, the lifetime $\tau_{\text{hetero}}(\tau_{\text{ngp}})$ continues to increase dramatically with decreasing temperature.
Thus, we can conclude that the lifetime $\tau_{\text{hetero}}$ gets large dramatically with decreasing temperature, whereas the correlation length $\xi_4$ and the intensity $\chi_4$ increase slowly compared to $\tau_{\text{hetero}}$ or plateau, i.e., the time scale of dynamical heterogeneity grows faster than the length scale and the intensity of dynamical heterogeneity.

\section{RESULTS II: TIME-INTERVAL DEPENDENCE OF LIFETIME} \label{result2}
In this section, we present the results for the time-interval dependence of the lifetime $\tau_{\text{hetero}}(t)$.
To examine the lifetime in more detail, we evaluated the lifetime $\tau_{\text{hetero}}(t)$ for various time intervals $t$ and determine how the lifetime $\tau_{\text{hetero}}(t)$ depends on the time interval.

\begin{figure}
\begin{center}
\includegraphics[scale=1]{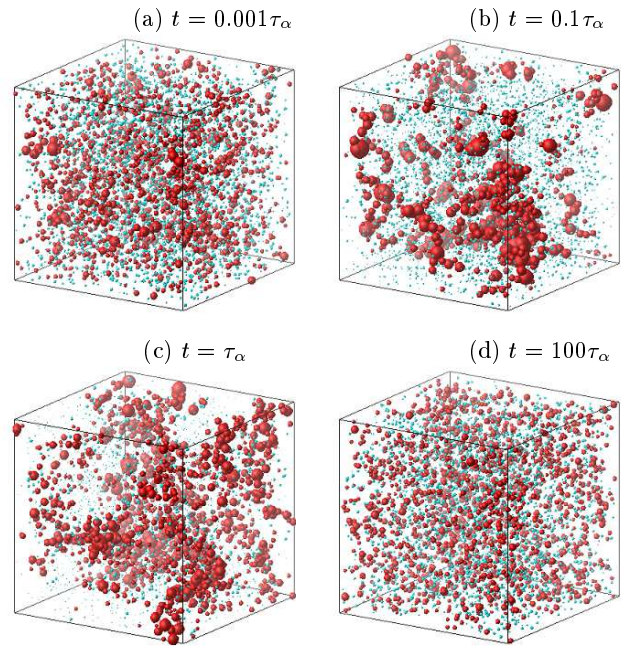}
\end{center}
\vspace*{-3mm}
\caption{(Color) 
The time-interval dependence of the distribution of the particle mobility $a_{1j}^2(t_0,t)$ for particle species 1.
The time intervals are (a) $t=0.001\tau_\alpha$, (b) $t=0.1\tau_\alpha$, (c) $t=\tau_\alpha$ and (d) $t=100\tau_\alpha$.
The temperature is $0.267$.
The radii of the spheres are $a_{1j}^2(t_0,t)$, and the centers are at $\vec{R}_{1j}(t_0,t)$.
See also the caption for Fig. \ref{hetero1}.
}
\label{hetero2}
\end{figure}

\begin{figure}
\begin{center}
\includegraphics[scale=1]{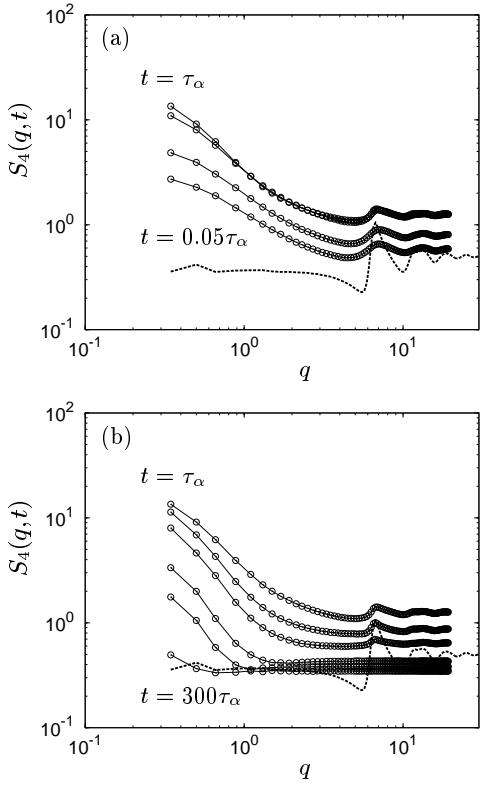}
\end{center}
\vspace*{-3mm}
\caption{The wavenumber dependence of $S_4(q,t)$ at $T=0.306$.
The time intervals are (a) $0.05\tau_\alpha$, $0.1\tau_\alpha$, $0.5\tau_\alpha$, and $\tau_\alpha$ from the lowest curve to the highest, and (b) $\tau_\alpha$, $5\tau_\alpha$, $10\tau_\alpha$, $50\tau_\alpha$, $100\tau_\alpha$, and $300\tau_\alpha$ from the highest curve to the lowest.
$S_4(q,t)$ is maximized at $t=\tau_\alpha$.
The dashed curve represents the static structure factor $S_{11}(q)$.
}
\label{sd4q}
\end{figure}

\subsection{The structure of the heterogeneous dynamics} \label{strct}
First, we examined the time-interval dependence of the structure of the heterogeneous dynamics.
The spatial distribution of the particle mobility $a_{1j}^2(t_0,t)$ for particle species 1 is shown in Fig. \ref{hetero2}, in which the time interval increases from $t=0.001 \tau_\alpha$ to $100\tau_\alpha$.
The temperature is $0.267$.
We can recognize that the heterogeneity is much weaker at the short time interval $t = 0.001 \tau_\alpha$ in \ref{hetero2}(a).
As the time interval $t$ increases, the heterogeneity increases and is maximized between $t = 0.1 \tau_\alpha$ in \ref{hetero2}(b) and $t=\tau_\alpha$ in \ref{hetero2}(c).
As the time interval $t$ increases further, the heterogeneity decreases and is weakened at $t = 100 \tau_\alpha$ in \ref{hetero2}(d).

We also calculated the spatial correlation function $S_{4}(q,t)$ defined in Eq. (\ref{sd4eq}).
Figure \ref{sd4q} shows the wavenumber dependence of $S_{4}(q,t)$ for various time intervals $t$.
We can see that as the time interval $t$ gets large, $S_{4}(q,t)$ increases, is maximized, and then decreases in the region of $q$.
This behavior agrees with the visualization of the heterogeneity structure shown in Fig. \ref{hetero2}.
It is also seen that at large time intervals, $S_{4}(q,t)$ maintains a constant value, independent of the wavenumber $q$.
This result indicates that at large time intervals, particle mobilities are uniformly distributed throughout space, i.e., the structure of the particle dynamics is spatially homogeneous.
Note that the static structure factor $S_{11}(q)$ also becomes constant at small $q$ (long-distance scale).

\subsection{The motion of the heterogeneous dynamics} \label{mhd}
Next, the motion of heterogeneous dynamics was investigated in detail.
In our previous study \cite{mizuno_2010}, we investigated the motion of the heterogeneous dynamics at the time intervals $t=\tau_\alpha$ and $\tau_{\text{ngp}}$.
Our results suggested that the heterogeneous dynamics might migrate in space with a diffusion-like mechanism.
In the present study, we examined the motion of the heterogeneous dynamics at various time intervals $t$.
As in our previous study \cite{mizuno_2010}, we calculated the time correlation function of the particle dynamics expressed in Eq. (\ref{F4}).
We used $\delta \vd_1(\vec{q},t_0,t)$ as $\delta Q_{{k}}(\vec{q},t_0,t)$ in Eq. (\ref{F4}), and the correlation function $S_{\vd}(q,t_s,t)$,
\begin{equation}
S_{\vd}(q,t_s,t) = \langle \delta \vd_1(\vec{q},t_s+t,t) \delta \vd_1(-\vec{q},0,t) \rangle,
\end{equation}
corresponds to $F_{4,{k}}({q},t_s,t)$ and represents the correlation of the particle dynamics between two time intervals $[0,t]$ and $[t_s+t,t_s+2t]$.
The time configuration of $S_{\vd}(q,t_s,t)$ is schematically illustrated in Fig. \ref{ts}(b).
As mentioned for Eq. (\ref{F42}), when the time separation $t_s$ increases, $S_{\vd}(q,t_s,t)$ with fixed $t$ decays in the stretched exponential form,
\begin{equation}
\begin{aligned}
\frac{S_{\vd}(q,t_s,t)}{S_{\vd}(q,0,t)}
\sim \exp\left( - \left( \frac{t_s}{\tau_{h}(q,t)} \right)^c \right),
\end{aligned}
\end{equation}
where $\tau_{h}(q,t)$ is the wavenumber-dependent relaxation time of $S_{\vd}(q,t_s,t)$ and represents the time scale at which the heterogeneous dynamics move in the space.
The expression $\tau_{h}(q,t)$ corresponds to $\tau_{4,{k}}({q},t)$ in Eq. (\ref{F42}).
We were able to examine the motion of the heterogeneous dynamics using the wavenumber dependence of $\tau_{h}(q,t)$.

\begin{figure}
\begin{center}
\includegraphics[scale=1]{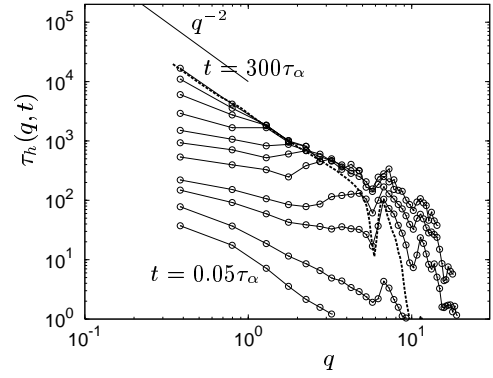}
\end{center}
\vspace*{-3mm}
\caption{The wavenumber dependence of $\tau_h(q,t)$ at $T=0.306$.
The time intervals are $0.05\tau_\alpha$, $0.1\tau_\alpha$, $0.5\tau_\alpha$, $\tau_\alpha$, $5\tau_\alpha$, $10\tau_\alpha$, $30\tau_\alpha$, $50\tau_\alpha$, $70\tau_\alpha$, $100\tau_\alpha$, and $300\tau_\alpha$ from the lowest curve to the highest.
The dashed curve is the relaxation time $\tau_{c1}(q)$ of the two-point density correlation function.
}
\label{tauhtsq306}
\end{figure}

In Fig. \ref{tauhtsq306}, we show the wavenumber dependence of $\tau_h(q,t)$ for various time intervals $t$.
As the time interval $t$ increases, at small wavenumbers $q$, $\tau_h(q,t)$ increases monotonically and approaches the relaxation time $\tau_{c1}(q)$ of the two-point density correlation function shown in Fig. \ref{tauk}.
This result indicates that the heterogeneous dynamics at large time intervals behave like the particle density, the motion of which is diffusive at a long distance scale.
Notice that $\tau_h(q,t)$ at large $t$ and small $q$ is proportional to $q^{-2}$, which indicates this diffusion mechanism.
We also checked that $S_{\vd}(q,t_s,t)$ decays in the exponential form at large $t$ and small $q$, as does the two-point density correlation function.

\begin{figure}
\begin{center}
\includegraphics[scale=1]{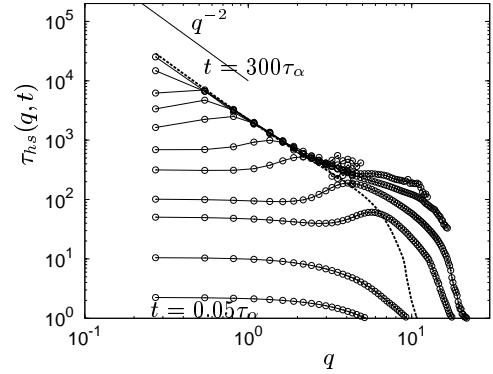}
\end{center}
\vspace*{-3mm}
\caption{The wavenumber dependence of $\tau_{hs}(q,t)$ at $T=0.306$.
The time intervals are $0.05\tau_\alpha$, $0.1\tau_\alpha$, $0.5\tau_\alpha$, $\tau_\alpha$, $5\tau_\alpha$, $10\tau_\alpha$, $30\tau_\alpha$, $50\tau_\alpha$, $70\tau_\alpha$, $100\tau_\alpha$, and $300\tau_\alpha$ from the lowest curve to the highest.
The dashed curve is the relaxation time $\tau_{s1}(q)$ of the self-part of the two-point density correlation function.
}
\label{tauhtssq306}
\end{figure}

Furthermore, we calculated the self-part of $S_{\vd}(q,t_s,t)$, defined by:
\begin{equation}
S_{\vd s}(q,t_s,t) = \left< \frac{1}{N_1} \sum_{j=1}^{N_1} \delta \vd_{1j}(\vec{q},t_s+t,t)\delta \vd_{1j}(\vec{q},0,t) \right>,
\end{equation}
where
\begin{equation}
\delta \vd_{1j}(\vec{q},t_0,t)= \delta a_{1j}^2(t_0,t) \exp[-i\vec{q} \cdot \vec{R}_{1j}(t_0,t) ].
\end{equation}
The correlation function $S_{\vd s}(q,t_s,t)$ represents the correlation of the individual particle dynamics between two time intervals $[0,t]$ and $[t_s+t,t_s+2t]$.
Like $S_{\vd}(q,t_s,t)$, as the time separation $t_s$ increases, $S_{\vd s}(q,t_s,t)$ decays in the stretched exponential form with the relaxation time $\tau_{hs}(q,t)$. The value of $\tau_{hs}(q,t)$ is the time scale at which the individual heterogeneous dynamics move in space. 
Figure \ref{tauhtssq306} shows the wavenumber dependence of $\tau_{hs}(q,t)$ for various time intervals.
As the time interval $t$ increases, $\tau_{hs}(q,t)$ increases monotonically at small $q$ and approaches the relaxation time $\tau_{s1}(q)$ of the self-part of the two-point density correlation function shown in Fig. \ref{tauk}.
Like $\tau_{h}(q,t)$, $\tau_{hs}(q,t)$ is also proportional to $q^{-2}$ at large $t$ and small $q$.
Thus, both the collective-particle behavior and the single-particle behavior of the heterogeneous dynamics at large time intervals are diffusive, like the motion of particle density at long-distance scales.

That $\tau_h(q,t)$ and $\tau_{hs}(q,t)$ approach the relaxation times $\tau_{c1}(q)$ and $\tau_{s1}(q)$ of the two-point density correlation functions can possibly be interpreted as follows.
The correlation function $S_{\vd}(q,t_s,t)$ can be written as
\begin{equation}
\begin{aligned}
& S_{\vd}(q,t_s,t) = \bigg< \sum_{j=1}^{N_1} \sum_{k=1}^{N_1} \delta a_{1j}^2(t_s+t,t) \delta a_{1k}^2(0,t) \times \\
& \qquad \quad \quad \exp[-i\vec{q} \cdot (\vec{R}_{1j}(t_s+t,t)-\vec{R}_{1k}(0,t)) ] \bigg>.
\end{aligned}
\end{equation}
From this expression, there can be two types of relaxation of $S_{\vd}(q,t_s,t)$ with increasing time of separation $t_s$.
One is the relaxation due to fluctuations in the particle mobility, i.e., due to the term ``$\delta a_{1j}^2(t_s+t,t) \delta a_{1k}^2(0,t)$''.
The other is the relaxation due to particle motion, i.e., due to the term ``$\exp[-i\vec{q} \cdot (\vec{R}_{1j}(t_s+t,t)-\vec{R}_{1k}(0,t)) ]$'', which is the same as the relaxation of the density correlation functions.
So, the total relaxation time $\tau_{h}(q,t)$ of $S_{\vd}(q,t_s,t)$ is determined by the two times $\tau_{\delta a}(t)$ and $\tau_{c1}(q)$, where $\tau_{\delta a}(t)$ is the time scale at which particle mobility fluctuates, and $\tau_{c1}(q)$ is the time scale of particle motion and is the relaxation time of the density correlation function.
Here, for simplicity, we assume that $S_{\vd}(q,t_s,t)$ relaxes with these two times $\tau_{\delta a}(t)$ and $\tau_{c1}(q)$ in the form
\begin{equation}
\begin{aligned}
& \frac{S_{\vd}(q,t_s,t)}{S_{\vd}(q,0,t)} \sim \exp\left( - \left( \frac{t_s}{\tau_{\delta a}(t)} \right)^c \right) \times \exp\left( - \left( \frac{t_s}{\tau_{c1}(q)} \right)^c \right).
\label{2times}
\end{aligned}
\end{equation}
The relaxation time $\tau_h(q,t)$ is then mainly determined by the smaller time scale of $\tau_{\delta a}(t)$ and $\tau_{c1}(q)$.
When the time interval $t$ is small, particle mobilities fluctuate faster than the particles move, which means that $\tau_{\delta a}(t)$ is smaller than $\tau_{c1}(q)$. In this case, $\tau_h(q,t)$ is determined by the time $\tau_{\delta a}(t)$.
As the time interval $t$ increases, then the fluctuations in particle mobilities becomes slower, and the lifetime $\tau_h(q,t)$ increases accordingly.
As the time interval $t$ increases further, particle mobilities fluctuate more slowly than the particles move, which means that $\tau_{\delta a}(t)$ becomes larger than $\tau_{c1}(q)$. In this case, $\tau_h(q,t)$ is determined by the time $\tau_{c1}(q)$ instead of $\tau_{\delta a}(t)$.
Thus, at large time intervals, because particle mobilities fluctuate very slowly, the relaxation time $\tau_h(q,t)$ can be determined by the time scale of particle motions, which is the relaxation time of the density correlation function.

\begin{figure}
\begin{center}
\includegraphics[scale=1]{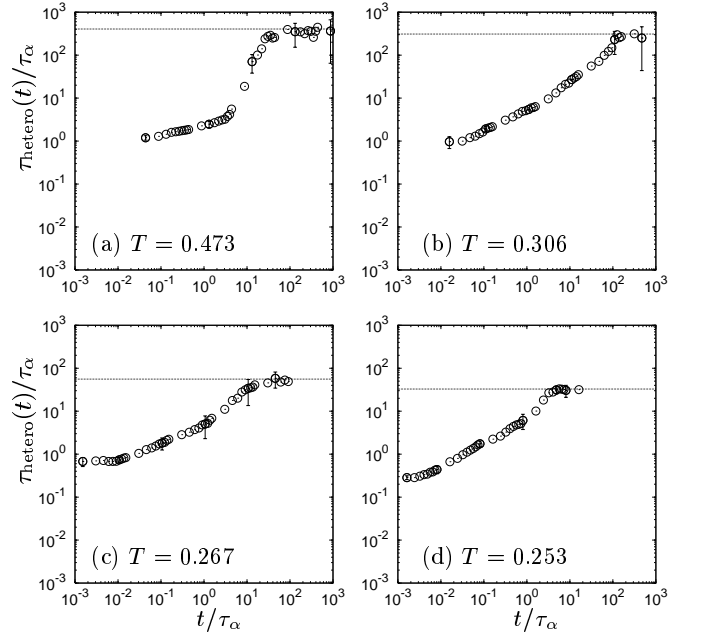}
\end{center}
\vspace*{-3mm}
\caption{The time-interval dependence of $\tau_{\text{hetero}}(t)$.
Temperatures are 0.473 in (a), 0.306 in (b), 0.267 in (c), and 0.253 in (d).
The time interval $t$ and the lifetime $\tau_{\text{hetero}}(t)$ are normalized by $\tau_\alpha$.
The dotted line indicates the value of $\tau_{c1}(q)$ at $q=0.38$, where $\tau_{c1}(q)$ is the relaxation time of the two-point density correlation function.
}
\label{tauhts}
\end{figure}

\subsection{The lifetime of the heterogeneous dynamics}
Finally, we determined the lifetime $\tau_{\text{hetero}}(t)$ of the heterogeneous dynamics as $\tau_{h}(q,t)$ at $q=0.38$.
We show the time-interval dependence of the lifetime $\tau_{\text{hetero}}(t)$ for various temperatures in Fig. \ref{tauhts}.
As the time interval $t$ increases, $\tau_{\text{hetero}}(t)$ increases monotonically, and at large time intervals, $\tau_{\text{hetero}}(t)$ approaches and is limited to the relaxation time $\tau_{c1}(q=0.38)$ of the two-point density correlation function.
At large time intervals for which $\tau_{\text{hetero}}(t)$ plateaus, the heterogeneous dynamics migrate in space with a diffusion mechanism like that of particle density, as we showed in Figs. (\ref{tauhtsq306}) and (\ref{tauhtssq306}).

\section{SUMMARY} \label{conclusion}
In this study, we have investigated three quantities that characterize dynamical heterogeneity: the correlation length $\xi_4(t)$, the intensity $\chi_4(t)$, and the lifetime $\tau_{\text{hetero}}(t)$.
The intensity $\chi_4(t)$ measures the average variance of the slow and fast regions, and the correlation length $\xi_4(t)$ characterizes the spatial extent of the slow and fast regions.
The lifetime $\tau_{\text{hetero}}(t)$ represents the time scale at which the slow and fast regions migrate in space.
We evaluated all three quantities using a single order parameter representing the particle dynamics and its correlation functions.
To define the particle dynamics, we used two time intervals, $t=\tau_\alpha$ and $\tau_{\text{ngp}}$.
We found that at low temperatures, the lifetime $\tau_{\text{hetero}}(t)$ increases dramatically with decreasing temperature. In contrast, the correlation length $\xi_4(t)$ and the intensity $\chi_4(t)$ increase slowly compared to the lifetime or plateau.
At the time interval $t=\tau_\alpha$, we obtained the scaling relationships between $\xi_4(t)$, $\chi_4(t)$, and $\tau_{\text{hetero}}(\tau_\alpha)$: $\chi_4(\tau_\alpha) \sim \xi_4(\tau_\alpha)^{3.2}$ and $\tau_{\text{hetero}}(\tau_\alpha) \sim \xi_4(\tau_\alpha)^{10.8}$ or $\tau_{\text{hetero}}(\tau_\alpha) \sim \exp(k \xi_4(\tau_\alpha)^{1.3})$.
The scaling exponent $10.8$ of $\tau_{\text{hetero}}(\tau_\alpha) \sim \xi_4(\tau_\alpha)^{10.8}$ is very large, and there is an exponential growth of $\tau_\text{hetero}(\tau_\alpha)$ with $\xi_4(\tau_\alpha)$.
Furthermore, at the time interval $t=\tau_\text{ngp}$, although the correlation length $\xi_4(\tau_{\text{ngp}})$ plateaus as the temperature decreases, the lifetime $\tau_{\text{hetero}}(\tau_{\text{ngp}})$ continues to increase dramatically.
Thus, we can conclude that the lifetime $\tau_{\text{hetero}}$ gets large dramatically with decreasing temperature, whereas the correlation length $\xi_4$ and the intensity $\chi_4$ increase slowly compared to $\tau_{\text{hetero}}$ or plateau, i.e., the time scale of dynamical heterogeneity grows faster than the length scale and the intensity of dynamical heterogeneity.

Furthermore, we investigated the time-interval dependence of the lifetime $\tau_{\text{hetero}}(t)$.
As the time interval $t$ increases, $\tau_{\text{hetero}}(t)$ increases monotonically.
At large time intervals, the lifetime $\tau_{\text{hetero}}(t)$ approaches and is limited to the relaxation time of the two-point density correlation function.
At those large time intervals, the wavenumber-dependent lifetimes $\tau_h(q,t)$ at small wavenumbers $q$ (long-distance scales) almost coincide with the relaxation time $\tau_{c1}(q)$ of the two-point density correlation function and are proportional to $q^{-2}$.
Therefore, the heterogeneous dynamics migrate in space with a diffusion mechanism like that of particle density.
Note that at large time intervals, particle mobilities $a_{1j}^2(t_0,t)$ are uniformly distributed in space, and the heterogeneity structure is much weaker.

\begin{acknowledgments}
We wish to acknowledge Dr. K. Kim for useful comments and discussions.
\end{acknowledgments}

\bibliography{paper}

\end{document}